\documentstyle[prl,aps,epsf]{revtex}
\begin{document}
\twocolumn[\hsize\textwidth\columnwidth\hsize\csname@twocolumnfalse\endcsname


\draft
\title{Effects of Disorder on the Competition between
Antiferromagnetism and Superconductivity}

\author{
Hiroshi Kohno\footnote{E-mail: kohno@watson.phys.s.u-tokyo.ac.jp},
Hidetoshi Fukuyama and Manfred Sigrist$^1$}
\address{
Department of Physics, University of Tokyo, Bunkyo-ku, Tokyo 113, Japan \\ 
$^1$ Yukawa Institute, Kyoto University, Kyoto 606-8502, JAPAN}
\date{\today}
\maketitle
\begin{abstract}
  Motivated by the observation of unusual magnetism in Ce$_x$Cu$_2$Si$_2$ 
($x\!\sim\! 1$), 
we study the effect of disorder, such as Ce vacancy, on the competition 
between superconductivity (SC) and antiferromagnetism (AF) 
on the basis of the phenomenological Ginzburg-Landau theory. 
  Assuming that the AF-SC transition is of first order in clean system, 
we show that a single impurity in the SC state 
can induce staggered magnetization by suppressing the SC around it. 
  For finite concentration of impurities, 
the first-order AF-SC boundary in the clean case is 
replaced by a finite region where the SC and the induced AF moments 
coexist microscopically with 
spatially varying order parameters.
  We argue that spin excitation spectrum in the coexistent state has a 
dual structure of SC gapped mode and the low-energy spin-wave mode. 
  In accordance with the emergence of AF out of SC ground state, 
the spectral weight will be transferred from the former mode to the latter,
keeping the structure of both modes basically unchanged.

\ 

\

\end{abstract}

]

\vskip 1cm
\section{Introduction}

  The interplay between magnetism and superconductivity (SC) is one of the 
interesting and profound phenomena ubiquitous in strongly correlated electron 
systems such as high-$T_{\rm c}$ cuprates, \lq low-$T_{\rm c}$' ruthenates, 
heavy-electron compounds and molecular solids. 
  Especially, heavy-electron systems often exhibit mysterious magnetism  
and exotic superconductivity, most of which remain to be characterized 
\cite{HF-review}.

  The heavy-electron compound
Ce$_x$Cu$_2$Si$_2$ $(x\!\sim\! 1)$ is located near the phase
boundary between superconductivity and antiferromagnetism (AF),
and shows an extreme sensitivity to the variation of the Ce concentration
$x$\cite{Steglich,Kitaoka,Nakamura,mu1,mu2,mu3,mu4,Ishida}. 
  Upon reducing $x$, the ground state is transformed from the spin-singlet 
SC ($x\!=\! 1.025, 1.00$) to AF ($x\!=\! 0.975$). 
  For $x\!=\! 0.99$, several experiments 
indicate the 
coexistence of SC and the so-called A-phase, the latter being characterized 
by the presence of spin fluctuations with unusually low frequency.
  Recent NQR measurements by Ishida {\it et al.} \cite{Ishida}
revealed that the energy scale of spin fluctuations in the A-phase
is extremely low and of the order of NQR frequency,
$\omega_{\rm NQR} \!\sim $1-10 MHz.
  Zero-field $\mu$SR studies also indicated the existence of random or 
slowly-fluctuating internal magnetic field\cite{mu1,mu2,mu3,mu4}.

  In this letter, we propose that in Ce$_x$Cu$_2$Si$_2$ with 
such a particular type of 
magnetism an essential role is played by disorder \cite{ISSP7}.
  A Ce vacancy, for example, can be the origin of such disorder, 
which will strongly disturb the electron coherence around it. 
  In the SC state, it will act as a pair breaker and 
locally reduce the SC amplitude. 
  It may even induce local magnetization, possibly the staggered one, 
around it. 
  We present a phenomenological description based on the 
Ginzburg-Landau (GL) theory for such effects of disorder on the SC state.  
  A key assumption made is that the SC state is in competition with the 
AF phase with a first-order phase boundary in clean case, 
which enables the AF phase to nucleate where SC is suppressed. 
  We limit the discussion to absolute zero, namely, the ground state.

  Disorder-induced AF in 
spin-gap systems has been recognized through studies of impurity effects 
in inorganic spin-Peierls system CuGeO$_3$ \cite{SP-exp,SP-th}. 
  In this system, a small amount of replacement of Ge by Si, or Cu by Zn 
induces local staggered moment with local reduction of lattice dimerization, 
leading to the coexistence of both long-range orderings (LRO's), 
lattice dimerization and AF, since there is no frustration for AF as
emphasized in ref.\cite{SP-th}.
  The present work attempts a general treatment for such phenomena,
especially in higher dimensions, using the GL theory. 
  Details will appear elsewhere.

\section{Clean Systems}

  We begin by considering a clean system, {\it i.e.}, without 
impurities, described by the GL free energy density 
\begin{eqnarray}
  F = &-& \alpha \psi^2 + \frac{\beta}{2} \psi^4
       - \alpha_M M^2 + \frac{\beta_M}{2} M^4 
       + \gamma \psi^2 M^2  \nonumber \\
      &+& \frac{\rho}{2} (\nabla \psi)^2 + \frac{\rho_M}{2} (\nabla M)^2. 
\label{F}
\end{eqnarray} 
  Both order parameters, $\psi$ and $M$, for SC and AF, respectively, 
are assumed to be real scalars\cite{com01}. 
  We are concerned only with their amplitude 
modulations here and neglect the phase (orientational) degrees of freedom for 
$\psi$ ($M$).   
  The constants, 
$\alpha, \beta, \alpha_M, \beta_M, \gamma, \rho$ and $\rho_M$, 
are all assumed to be positive. 
  Especially, we consider the case where the ground state is always in a 
broken-symmetry phase of either SC or AF, which is expressed by 
the conditions, $\alpha >0$ and $\alpha_M >0$. 
  This viewpoint has been taken by Zhang in the $SO(5)$ theory \cite{so5}.   
  We also assume that there is no coexistence of SC and AF in the 
clean case, which is ensured by the condition, 
$ \gamma' > {\rm min} \{ \beta', \beta_M'\} $, 
where 
$ \gamma' = \gamma / \alpha \alpha_M $, 
$ \beta'  = \beta / \alpha^2 $ and 
$ \beta_M'  = \beta_M / \alpha_M^2 $. 
  Then the ground state is 
SC if $ \beta' < \beta_M' $ and AF otherwise.

  In the following, we consider the parameter region where the SC 
state is stabilized in the bulk clean system, but is located near the 
phase boundary to the AF state. 
  The SC order parameter takes a constant value 
$\psi_0 \equiv \sqrt{\alpha / \beta}$ in the bulk in this case.

\section{Single Impurity}

  We now introduce a single impurity (at the origin) into the 
otherwise uniform SC ground state. 
  For CeCu$_2$Si$_2$, this 
may represent a Ce vacancy, for example. 
  Assuming that the impurity acts as a local pair breaker for SC and 
locally favors AF relative to SC, 
we model its effect by
\begin{eqnarray}
  \alpha  \ &\longrightarrow& \ 
  \alpha ({\bf r}) \equiv \alpha \bigl[ 1 - V_0 \delta ({\bf r}) \bigr], \\
  \alpha_M  \ &\longrightarrow& \ 
  \alpha_M ({\bf r}) \equiv \alpha_M \bigl[ 1 + V_1 \delta ({\bf r}) \bigr], 
\label{imp-a0}
\end{eqnarray}
where $V_0, V_1 \geq 0$.  
  The GL equations are given by 
\begin{eqnarray}
 \frac{\rho}{2} \nabla^2 \psi ({\bf r})
    &=& \bigl[ - \alpha ({\bf r}) + \gamma M^2 \bigr] \psi + \beta \psi^3, 
\label{GL-eqD0} \\
  \frac{\rho_M}{2} \nabla^2 M ({\bf r})
    &=& \bigl[ - \alpha_M ({\bf r}) + \gamma \psi^2 \bigr] M 
       + \beta_M M^3. 
\label{GL-eqM0} 
\end{eqnarray}

  Equation (\ref{GL-eqD0}) describes the suppression of the SC order parameter 
$\psi$ around the impurity. 
  If we neglect the $\gamma$ term for simplicity\cite{com02}, 
the solution is written as
\begin{equation}
 \psi ({\bf r}) = \sqrt{\alpha /\beta } \, f(r; \xi, V_0)
\label{psi}
\end{equation}
with $\xi = \sqrt{\rho / 2 \alpha }$ and $r = |{\bf r}|$.
  Qualitative features of the function $f$ are as follows.
  It is suppressed appreciably from the bulk value of 1 only within a
region $r < \kern -11.7pt \lower 4.3pt \hbox{$\displaystyle \sim$} \xi $, 
where it decreases almost linearly with $r$ as $ r \rightarrow 0$.
  The value $f(r\!=\! 0)$ at the origin is a monotonically decreasing
function of $V_0$;
it is 1 for $V_0 = 0$ and vanishes as $V_0 \rightarrow \infty$.

  Equation (\ref{GL-eqM0}) describes the possible emergence of staggered 
moment $M$ around the impurity.  
  To study this, we first drop the non-linear term ($\sim M^3$) 
and consider the following eigenvalue problem
\begin{equation}
  \biggl[ -\frac{\rho_M}{2} \nabla^2 + U_M ({\bf r}) \biggr] M ({\bf r})
  = \varepsilon M ({\bf r}), 
\label{GL-eqM1}
\end{equation}
where the \lq potential' $U_M ({\bf r})$ is given by
\begin{equation}
  U_M ({\bf r}) = \tilde \alpha_M 
                 - \gamma \bigl[ \psi_0^2 - \psi^2({\bf r}) \bigr] 
                 - \alpha_M V_1 \delta ({\bf r}). 
\label{GL-eqM2}
\end{equation}
  In eq.(\ref{GL-eqM2}), the first term, 
$ \tilde \alpha_M \equiv - \alpha_M + \gamma \psi_0^2 $, 
is positive\cite{com08}, which  
assures the local stability (\lq local' in the $\psi$-$M$ space) 
of the pure SC state in the clean case. 
  The second and the third terms represent \lq attractive potential' 
due to the suppression of $\psi$, and the direct coupling to the impurity, 
respectively. 
  If we have a 
non-zero 
solution $M ({\bf r})$ with negative \lq energy' 
$\varepsilon$, this 
signals the appearance or \lq nucleation' of the AF 
phase around the impurity in the otherwise SC system\cite{com09}.
  Its spatial pattern is given by the corresponding eigenfunction
(with lowest eigenvalue, $\varepsilon$) and the amplitude is determined 
by balancing the quadratic term ($\sim\! \varepsilon M^2$) with the 
quartic term ($\sim\! \beta_M M^4$),  
which leads to $ M \sim \sqrt{-\varepsilon / \beta_M} $.

  As we have seen, the characteristic length-scale for $\psi$ is given by
$\xi$. 
  On the other hand, spatial variation of $M$ occurs on the scale, 
$ \xi_M = \sqrt{\rho_M / 2 \alpha_M}$.\cite{com1}
  Depending on their ratio, $ \xi/\xi_M $, 
there are two different regimes, namely, 
$\xi \ll \xi_M$ (type I) and $\xi \gg \xi_M$ (type II).  
  The intermediate regime ($\xi \!\sim\! \xi_M$) may be the case 
where the $SO(5)$ symmetry, 
a rotational symmetry in the $\psi$-$M$ space, is expected. 
  We examine these three cases in the following.

  {\bf (i) Type I} ($\xi \ll \xi_M$): 
  The suppression of $\psi$ is limited to a small region compared 
to $\xi_M$.
  Its effect on $M$ can be shown to be negligible, 
and the $V_1$-term is necessary for $M$ to nucleate. 
  Rewriting eq.(\ref{GL-eqM1}) as 
\begin{equation}
  (-\nabla^2 + \kappa_M^2) M ({\bf r}) = v_1 M_0 \delta ({\bf r}), 
\label{GL-eqM3} 
\end{equation}
with
$ v_1 = V_1/\xi_M^2 $, 
$ \kappa_M = \kappa_M (\varepsilon )
   \equiv \sqrt{2(\tilde \alpha_M - \varepsilon)/\rho_M}$,
and 
$  M_0 \equiv M({\bf r}\!=\! {\bf 0})$, 
the solution is obtained as
\begin{equation}
  M({\bf r}) = v_1 M_0 G({\bf r}; \kappa_M), 
\label{M1} 
\end{equation}
where 
$ G({\bf r};\kappa) $
is the Green's function satisfying 
$ (-\nabla^2 + \kappa^2 )  G({\bf r};\kappa) = \delta ({\bf r})$. 
  The explicit form is given by 
\begin{equation}
  G({\bf r};\kappa)
 = {\rm e}^{-\kappa r}/2\kappa, \ \ 
   K_0(\kappa r)/2\pi,          \ \ 
   {\rm e}^{-\kappa r}/4\pi r, 
\label{G}
\end{equation}
for dimensions $d=1,2$ and 3, respectively, 
where $K_0$ is the modified Bessel function of rank zero.  
  (The present Ce-compound corresponds to $d=3$.) 
  Equation (\ref{M1}) expresses a nucleated staggered magnetization 
if the solution exists such that $M_0 \ne 0$ 
and $\varepsilon \!<\! 0$.

  The condition for $M_0 \!\ne\! 0$ is obtained from (\ref{M1}) by putting 
${\bf r}={\bf 0}$ and dividing both sides by $M_0$ as
\begin{eqnarray}
  1 = v_1 G (r\!=\! a ; \kappa_M(\varepsilon )).
\label{SCF1}
\end{eqnarray}
  This is an eigenvalue equation for $\varepsilon$. 
  For $d=2$ and 3, the right hand side is evaluated by introducing a 
cut-off $a$ 
which will be of the order of the lattice constant\cite{com3}. 
  The nucleation of $M$ occurs for 
$ v_1 > v_{1, {\rm cr}}$ where 
\begin{equation}
  v_{1, {\rm cr}}
 = 2 / \tilde \xi_M, \ \
   2 \pi / K_0(a/\tilde \xi_M),          \ \
   4\pi a  \, {\rm exp} \bigl( a / \tilde \xi_M \bigr), 
\label{vcr}
\end{equation}
for $d=1,2$ and 3, respectively, 
with $ \tilde \xi_M = \sqrt{\rho_M / 2\tilde \alpha_M }$.  

  {\bf (ii) Type II} ($\xi \gg \xi_M$): 
  In this case, $\psi$ is suppressed in a wider region compared to $\xi_M$, 
and it is possible for $M$ to nucleate even without the $V_1$-term.
  As an example we consider the case $ V_0 \rightarrow \infty $ and 
$V_1=0$. 
Then the potential for $M$ is given by a harmonic one generated by the linear 
$r$-dependence of $\psi ({\bf r})$ around the impurity. 
  If we assume such harmonic potential in the entire space, the condition
for the nucleation of $M$ is obtained as $\xi_M < \xi /d^2$.

  {\bf (iii) \lq $SO(5)$ symmetric' case} ($\xi \!=\! \xi_M$):
  In case both length-scales are comparable, we must solve the coupled 
equations for $\psi$ and $M$ simultaneously. 
  However, a simplification is possible if the \lq anisotropy' 
is weak in the $\psi$-$M$ space. 
  We define the case what may be called \lq $SO(5)$ symmetric' 
by $ \xi = \xi_M $. 
  We then introduce the angular ($\theta$) and radial ($R$) variables by
$ (\alpha^{1/2} \psi, \alpha_M^{1/2} M) 
 = (R \cos \theta, R \sin \theta )$, 
to obtain 
\begin{equation}
  F = \xi^2 
      \bigl[ (\nabla R)^2 + R^2 (\nabla \theta )^2 \bigr] 
      -  R^2 + \frac{1}{2} \beta (\theta ) R^4   
\end{equation}
for clean systems, where 
$ \beta (\theta )  = \beta_0 + g \cos 2\theta + B \sin^2 2\theta $, 
$ \beta_0 = \textstyle{\frac{1}{2}} (\beta' + \beta'_M)$,
$ g = \textstyle{\frac{1}{2}} (\beta' - \beta'_M)$, and 
$ B = \textstyle{\frac{1}{2}} (\gamma' - \beta_0) $.
  The terms with $g$ and $B$ represent, respectively,
the \lq anisotropy energy' and the energy barrier between SC and AF.
  ($g<0$ favors SC relative to AF.)
  The condition of weak anisotropy, $ |g|, |B| \ll \beta_0 $, 
restricts the low-energy configurations to the manifold 
$ R = R(\theta ) \equiv \beta (\theta )^{-1/2}$, 
which as a function of $\theta$ 
has small variation compared to the average 
$ \sim R_0 \equiv \beta_0^{-1/2} $. 
  Introducing the impurity potential by 
$ F_{\rm imp} = V R^2 \cos 2\theta \delta ({\bf r})$, 
and considering the spatial variation of the order parameters, 
we again recognize two length-scales, 
$ \xi_R \sim \xi $ and 
$ \xi_\theta \sim \xi \sqrt{ \beta_0 / {\rm max}\{|g|,|B| \} }$, 
now associated with variables $R$ and $\theta$, respectively.
  From the above condition, we have $\xi_R \ll \xi_\theta$.
  Therefore, while both $R$ and $\theta$ deviates from the bulk values 
in the vicinity of the impurity ($r < \xi_R$), 
for $r>\xi_R$, only $\theta$ governs the spatial variation
and $R$ follows $\theta$ completely as $ R = R(\theta )$ given above. 
  The behaviours in the region $r>\xi_R$ will hence be described by the 
effective free energy, $F_\theta$: 
\begin{equation}
  F_\theta /R_0^2 = \xi^2 (\nabla \theta )^2 
      + g ({\bf r}) \cos 2\theta + B' \sin^2 2\theta,  
\end{equation}
where 
$ g ({\bf r}) = (g/2\beta_0) + V \delta ({\bf r})$, 
$ B' = B/2\beta_0$. 
  For $B=0$ and $d=1$, the present problem is equivalent to that of 
the disordered spin-Peierls system\cite{SP-th,SP-th2}. 
 The one with $B=0$ and $d=2$ (with $V=0$) corresponds to the $SO(5)$ 
description proposed for high-$T_{\rm c}$ cuprates\cite{so5}.

\section{Finite Concentration of Impurities}
  
  In the presence of many impurities (located at $\{{\bf R}_i\}$), 
the solution is written as
\begin{equation}
  {\bf M}({\bf r}) =  \sum_i {\bf m}_i M_1 ({\bf r}-{\bf R}_i) 
\label{M2}
\end{equation}
in the dilute concentration limit. 
  Here $M_1$ is the solution for the single-impurity problem 
obtained in the previous section, and 
we have recovered the vector nature of the AF order parameter. 
  The unit vector, ${\bf m}_i$, represents the direction of 
the staggered moment nucleated 
around the impurity at ${\bf R}_i$, and the interaction among them 
will be described by 
\begin{equation}
  F_{\rm m} = - \sum_{i < j} J_{ij} {\bf m}_i \!\cdot\! {\bf m}_j.  
\label{Fm}
\end{equation}
  The effective exchange constants 
\begin{equation}
  J_{ij} = (-2\varepsilon) \int {\rm d}{\bf r} 
           M_1({\bf r} - {\bf R}_i) M_1({\bf r} - {\bf R}_j) + \ldots 
\label{Heff}
\end{equation}
do not have frustration 
in the present model focussing on the commensurate AF, but will allow it 
in general. 
  For depleted quantum spin systems, similar models have been obtained 
previously\cite{Heff}.

  Orientational fluctuations of ${\bf m}_i$'s will produce low-energy 
spin excitations even when the LRO of SC 
is not completely destroyed by the emergence of $M$. 
  If ${\bf m}_i$'s align and exhibit AFLRO, their fluctuations will 
show up as a gapless Goldstone (spin-wave) mode.
  We note that the coexistence of both types of LRO, SC and 
impurity-induced AF, 
is possible due to the spatial variation of the two order parameters 
in spite of their competition.
  This general aspect has been first noticed in studies on disordered 
spin-Peierls systems.\cite{SP-th} 
  If AFLRO failed to be established for some reasons (see below), 
fluctuations of ${\bf m}_i$'s will remain as a diffusive mode 
with very low energy with the total spectral weight being proportional to 
$M_1^2$ and the concentration of impurities.

\section{Discussion}

  We have given an analysis based on the GL theory of the effect of 
disorder in the SC ground state which is in competition with the AF 
state.  
  Our interest is focussed on the nucleation
of the AF phase around disorder centers. 
  As the concentration of impurities is increased, the original SC state 
will first change into the microscopically coexistent state with 
nucleated staggered moments, 
and finally be transformed into the complete AF.  
  The first-order AF-SC boundary in the clean
case thus turns into a finite region of coexistence.

\begin{figure*}[htb]
\centerline{\epsfxsize=78mm \epsfbox{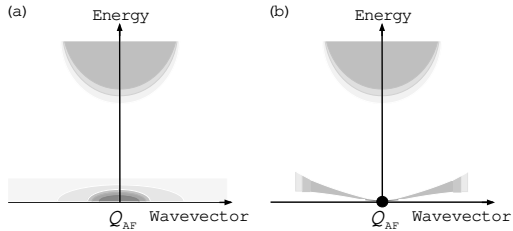}}
\caption{Spin excitation spectrum in the coexistent
phase of SC and disorder-induced staggered magnetization.
  It has a dual structure of SC gapped mode from quasiparticles
and the low-energy mode from nucleated staggered moments.
(a) In the absence of AFLRO, low-energy excitations are diffusive.
(b) In the presence of AFLRO, low-energy excitations are Goldstone
(spin wave) modes.  The dot at $Q_{\rm AF}$ indicates the Bragg spot of AF.
The energy scale for the low-energy mode is exaggerated.}
\label{FIG1}
\end{figure*}

  Analogously to disordered spin-Peierls systems\cite{SP-exp1,SP-th1}, 
the present coexistent state has spin excitations
with two distinct features, the SC gapped excitation due to 
quasiparticles and the gapless spin-wave mode, well separated from each 
other by a \lq transparent' region (see Fig.\ref{FIG1}). 
  In accordance with the emergence of AF out of SC ground state,
the spin excitation spectrum
will change mainly through the transfer of the spectral weight
from the gapped mode to the spin-wave mode, keeping the structure and
energy scales of both modes essentially unchanged, with a possible reduction 
of the SC gap and 
an increase of the spin-wave velocity.

  We would like to propose that the above picture describes well the 
evolution of the ground state from SC to AF in Ce$_x$Cu$_2$Si$_2$. 
  Especially, we propose that the A-phase realized near the phase boundary 
is characterized by a coexistence phase of SC and the nucleated staggered 
moments. 
  Experiments so far performed on samples in the A-phase did not detect 
any signals of magnetic LRO\cite{com4}, 
but revealed the existence of dynamical fluctuations 
with very low frequency of NQR scale\cite{Ishida}.
  Such low-energy fluctuations will be naturally understood as 
orientational fluctuations of the nucleated staggered moments around 
impurities. 
  In the absence of true AF LRO, they behave as diffusive (Fig.1(a)). 
  The spectral weight can be substantial at low energy  
since they are transformed into a Bragg spot and collective spin-wave 
modes as AF LRO is reached (Fig.1(b)).

  The nucleation of staggered moments will occur at some finite temperature. 
  Then what is their fate in the low-temperature limit where thermal 
fluctuations are diminished?
  There are several possibilities depending on the interaction $J_{ij}$ 
among them; 
(i) Constructive interaction (without frustration) leads to AFLRO, 
(ii) the presence of frustration leads to a spin-glass state, 
(iii) even if there is no frustration, the interaction among them
is so weak and their orientations freeze almost randomly and  
independently, resulting in a spin-glass-like state, and  
(iv) they continue fluctuating due to quantum effects.
  The spin-Peierls system Cu$_{1-x}$Zn$_x$GeO$_3$ offers an example of
case (i), where AF LRO has been observed in samples with $x$ being
as low as $1.12\times 10^{-3}$ with N${\rm \acute{e}}$el temperature
$T_N = 0.0285$K.\cite{Manabe}
  For Ce$_x$Cu$_2$Si$_2$ with $x=0.99$, the NQR experiment has been done 
down to temperature $ T = 0.012$K, which suggests the case (ii) or (iii). 
  If the nucleated staggered moments are incommensurate\cite{IC}, 
their interaction will be frustrated, leading to the case (ii).

  Our present picture predicts the spectral weight transfer between the 
\lq rigid bands' of gapped mode and the \lq spin-wave' mode (Goldstone  
or diffusive, depending on the presence or absence of AF LRO) 
as the ground state is tuned from SC to AF through the coexistent phase.  
  It will be interesting to explore experimentally  
this spectral-weight transfer by sweeping through the phase transition 
via, for example, pressure as proposed in ref.\cite{Ishida}. 

  In the coexistent state proposed in this paper, 
SC and AF are in competition even though they \lq coexist'. 
  The $\mu$SR studies\cite{mu2,mu3} seem to have revealed their competing 
nature experimentally, but then concluded a spatial seggregation of the two 
phases. 
  Before deriving such a conclusion, 
an analysis from the present viewpoint may be necessary.

  In conclusion, we have examined a possible nucleation of the AF 
phase due to disorder in the otherwise pure SC state and proposed that 
such a coexistent phase is responsible for the peculiar magnetism 
in the so-called A-phase of Ce$_x$Cu$_2$Si$_2$. 
  Such effects of disorder on the interplay between magnetic 
ordering and superconductivity (or spin gap in general) will be quite 
ubiquitous in strongly correlated systems and deserve attention 
in interpreting experimental data.

  We would like to thank Professor Y.~Kitaoka for 
a nice introduction to Ce$_x$Cu$_2$Si$_2$, as well as 
for valuable comments.


\begin{thebibliography}{99}
\bibitem{com01}
  We assume spin-singlet pairing (with one-dimensional representation) 
for SC, and commensurate structure for AF.
\bibitem{com02}
  The effect of the $\gamma$ term on $\psi$ can be taken into 
account as follows.
  For the type I case (defined below), we substitute 
$\alpha - \gamma M^2({\bf r})$ for $\alpha$
in the sense of an \lq adiabatic' approximation.
  For the type II case, it may be modeled  by the $\delta$-function 
potential for $\psi$, which just renormalizes $V_0$.
\bibitem{com09}
  The appearance of a bound state is not a sufficient condition
since there is a positive offset $\tilde \alpha_M$ to the \lq energy'
$\varepsilon$.
  A bound state with positive \lq energy' $\varepsilon$ is an excited
state.
\bibitem{com1} 
  Actually, however, $\xi$ and $\xi_M$ depend on the state, 
or the values of the order parameters. 
  For example, in a clean SC state where $\psi \equiv \psi_0$ and 
$M \equiv 0$, they are given by
$ \sqrt{\rho / 4 \alpha} $ and
$ \tilde \xi_M = \sqrt{\rho_M / 2 \tilde \alpha_M} $, respectively.
  In case where the order parameters have large spatial variation 
(which is our present concern), 
$\xi$ and $\xi_M$ should be obtained after solving the GL equations.
  In fact, the length scale for $M$ is given by
$\kappa_M^{-1} (\varepsilon )$ defined below.
  However, the following arguments remain valid.
\bibitem{com3}
  One way to see this is to consider the fourth-order term in gradient,
$ F \sim \rho_M (\nabla M)^2 + \rho_{M4} (\nabla^2 M)^2 $; then we have 
$a \sim \sqrt{\rho_{M4}/\rho_M}$.
\bibitem{IC} 
  The incommensurate nature of the magnetic ordering in 
Ce$_{0.975}$Cu$_2$Si$_2$ has been indicated experimentally 
in ref\cite{Ishida}.
\end{thebibliography}
\end{document}